# Detecting false correlations:

# Uncovering a faked Bell-inequality violation

M. E. Feldman,[1] G. K. Juul,[1] S. J. van Enk,[2] and M. Beck[1,]*

[1]*Department of Physics, Whitman College, Walla Walla, Washington 99362, USA*

[2]*Department of Physics, University of Oregon, Eugene, Oregon 97403, USA*

(Dated 7 February 2018)

It is possible for two parties, Alice and Bob, to establish a secure communication link by sharing an ensemble of entangled particles, and then using these particles to generate a secret key. One way to establish that the particles are indeed entangled is to verify that they violate a Bell inequality. However, it might be the case that Bob is not trustworthy and wishes Alice to believe that their communications are secure, when in fact they are not. He can do this by managing to have prior knowledge of Alice's measurement device settings and then modifying his own settings based upon this information. In this case it is possible for shared particle states that must satisfy a Bell inequality to *appear* to violate this inequality, which would also make the system appear secure. When Bob modifies his measurement settings, however, he produces false correlations. Here we demonstrate experimentally that Alice can detect these false correlations, and uncover Bob's trickery, by using loop-state-preparation-and-measurement (SPAM) tomography. More generally, we demonstrate that loop SPAM tomography can detect false correlations (correlated errors) in a two-qubit system without needing to know anything about the prepared states or the measurements, other than the dimensions of the operators that describe them.

## I. INTRODUCTION

It is well known that quantum physics allows, at least in principle, two parties (call them Alice and Bob) to create a shared secret key that they can use to encode a message that cannot be read by any eavesdropper. This is known as quantum-key distribution (QKD), and practical QKD systems now exist. While real systems are not perfectly secure, they allow Alice and Bob to obtain a very high degree of security and to know exactly how secure their key is [1].

One technique for Alice and Bob to secure their communications is to perform QKD using an ensemble of entangled particles. To demonstrate that their key is secure they can verify that their measurements satisfy a Bell inequality; this is the basis of device independent QKD [2, 3]. However, it is possible to exploit experimental loopholes, and "fake" a Bell inequality violation [4]. For example, if Bob has information about Alice's detector settings (e.g., he hacks into her computer) he can modify his own settings to make her believe that their particles violate a Bell inequality, even if the particles they exchange are in a state that cannot properly violate such an inequality. In this case the particles *appear* to violate local realism, but only because Bob has effectively introduced nonlocality into the system via his knowledge of Alice's detector settings. Bob may wish to fool Alice in this manner, for example, to allow his confidant to intercept their communications, unbeknownst to Alice. Indeed, it need not be Bob who is at fault. An eavesdropper who obtains prior information about Alice's and Bob's settings could perform select rotations on their particles, thus effectively changing their measurements. This would allow the eavesdropper to fool both Alice and Bob, and ultimately reduce the secrecy of their key. For simplicity, in what follows we will assume that it is Bob who is untrustworthy.

When Bob makes use of his knowledge of Alice's detector settings to modify his own settings he produces false correlations (also known as correlated errors): His measurement settings are now correlated with Alice's, not independent of them as would be the case in a true Bell-inequality test. Loop state-preparation-and-measurement (SPAM) tomography is a technique for detecting such false correlations, and Alice can use it to detect the correlations that Bob introduces with his fakery [5-8].

Loop SPAM tomography is useful for this purpose because it can detect correlated errors between state preparations and measurements, or between measurements performed at different locations, even without specific knowledge of any of the state preparations or the measurement settings [5-8]. The only assumption needed is that we know the dimensions of the Hilbert spaces that describe the state of the system (via its density operator), and the measurements (via the positive-value-operator measures, POVMs, that describe the detectors). By analyzing an over-complete set of measurements, loop SPAM tomography looks for self-consistency within the data. Here self-consistency means that for each state preparation we can assign a single state, and for each measurement we can assign a single POVM. If the data is not self-consistent in this manner, we say that there are false correlations. Hence, Alice can use loop SPAM tomography to detect the false correlations that Bob introduces because she doesn't need to know the state of their particles, and she doesn't need to trust what Bob tells her about his measurement settings.

Loop SPAM tomography has recently been demonstrated to be effective for detecting correlated errors between state preparations and measurements in a single-qubit system [8]. Here we show that it also works for detecting false correlations between different measurements performed on a two-qubit system. Beyond the particular experiments described here, our results demonstrate the ability to detect correlated errors in a device-independent way, with a minimum of assumptions, in a multiple qubit system. As such, loop SPAM tomography is useful for characterizing quantum information processors. Fault-tolerant quantum computers are susceptible to correlated errors, and as the fidelity of quantum operations improve they will be sensitive to ever smaller errors. It will be important to have a means of detecting these small but correlated errors, in order to improve our trust in the results of quantum computations.

To demonstrate the effectiveness of loop SPAM tomography for characterizing two-qubit systems we prepare photon pairs in states that approximate Werner states; these states are a mixture of a Bell state with a maximally mixed state (an incoherent background) [9]. The states we create have sufficient degree of mixture to ensure they are not capable of violating the Clauser-Horne-Shimony-Holt (CHSH) inequality $S \leq 2$ [10, 11]. However, Bob cheats by managing to have prior knowledge of Alice's detector settings, and he modifies his settings based on this knowledge in order to create measurements that yield $S > 2$. Yet all is not lost—Alice gets her revenge in the end, as she uses loop SPAM tomography in order to catch Bob's fakery.

In Sec. II we present a brief description of the theory of loop SPAM tomography. We describe our experimental setup and present our results in Sec. III, and in Sec. IV we have some concluding remarks.

## II. THEORY

### A. Loop SPAM Tomography

An unbiased detector that measures polarization is described by a POVM that has $n = 3$ independent parameters, and we can construct Hermitian operators (linear combinations of POVMs) that correspond to polarization measurements that also are described by three independent parameters [5, 8, 12]. Let Alice's measurement operators be denoted by $\hat{A}_i$, where $i$ labels the different measurement settings that Alice can choose. Similarly denote Bob's measurement operators by $\hat{B}_j$. If Alice and Bob have a source of photon pairs whose joint polarization state is described by the density operator $\hat{\rho}$, the expectation value for a joint measurement is given by

$$E_{ij} = \text{Tr}\left[\hat{\rho}\left(\hat{A}_i \otimes \hat{B}_j\right)\right]. \tag{1}$$

Assume that $\hat{\rho}$ is held constant during the measurements, and consider the $E_{ij}$'s to be elements of a matrix $\overline{E}$; we use an overbar to denote a quantity expressed as a matrix. Each row of $\overline{E}$ refers to a fixed measurement for Alice $\hat{A}_i$, while each column refers to a fixed measurement for Bob $\hat{B}_j$. Because each POVM has 3 independent parameters, the largest matrix $\overline{E}$ that can consist entirely of independent elements is 3x3.

Consider the case where Alice and Bob make an over-complete set of measurements. Alice makes measurements with $2n = 6$ different detector settings and so does Bob. The 6x6 matrix of expectation values $\overline{E}$ can be partitioned into corners consisting of 3x3 matrices as follows

$$\overline{E} = \begin{pmatrix} \overline{A} & \overline{B} \\ \overline{C} & \overline{D} \end{pmatrix}. \tag{2}$$

The $n$x$n$ corner matrix $\overline{A}$ consists of independent measurements, but the other corners cannot be independent of $\overline{A}$. For example, matrix $\overline{A}$ is connected to matrix $\overline{B}$ in the sense that they share a common set of Alice's measurements, and the measured matrix elements of $\overline{B}$ must be consistent with that fact.

Define the partial determinant of $\overline{E}$ as $\Delta(\overline{E}) \equiv \overline{A}^{-1}\overline{B}\,\overline{D}^{-1}\overline{C}$. It can be shown that the measured data are internally consistent as described above, and free of correlated SPAM errors, if and only if $\Delta(\bar{E}) = \bar{1}$, where $\bar{1}$ is the 3x3 identity matrix [5-8]. No knowledge of the state $\hat{\rho}$ or the detector operators is necessary to make this determination. Indeed, we don't even need to estimate the operators that describe the state or the detectors. The only assumption that we need to make is that we know the Hilbert-space dimensions of the state and the POVMs that describe the detectors.

If Bob has prior knowledge of Alice's detector settings, and uses that knowledge to change his detector settings to achieve a desired outcome, he will create a false correlation. Thus, to determine if Bob cheats we construct the matrix of expectation values $\bar{E}$ as given in Eq. (2), and then calculate the partial determinant $\Delta(\bar{E})$. If $\Delta(\bar{E}) - \bar{1} \neq 0$, to within the statistical errors of the measurements, then Bob's fakery has been detected. Furthermore, in order to detect Bob's cheating it is not necessary to use the full $2n = 6$ settings. Alice and Bob can each make $n + 1 = 4$ measurements, and embed these measurements into a 6x6 matrix $\bar{E}$ and test the condition $\Delta(\bar{E}) = \bar{1}$ [5, 7, 8].

**B. Horodecki *M* parameter**

We wish to determine whether or not the state that we create is capable of violating the CHSH inequality or not, and we use the method described in Ref. [11] to do this. We create the 3x3 matrix $\bar{T}$, whose elements are given by $T_{ij} = \text{Tr}\left[\hat{\rho}(\hat{\sigma}_i \otimes \hat{\sigma}_j)\right]$, where the $\hat{\sigma}_i$'s are the Pauli matrices. The symmetric matrix $\bar{U}$ is given by $\bar{U} = \bar{T}^T \bar{T}$, where $\bar{T}^T$ is the transpose of $\bar{T}$, and the two largest, positive, eigenvalues of $\bar{U}$ are denoted by $u$ and $\tilde{u}$. We define the parameter $M$ as $M \equiv u + \tilde{u}$; it was proven by Horodecki *et al* that $S_{\max} = 2\sqrt{M}$. Hence, $\hat{\rho}$ is capable of violating the CHSH inequality if and only if $M > 1$ [11].

## III. EXPERIMENTS

### A. Experimental Design

As shown in Fig. 1, we use a 150 mW, 405 nm laser diode to pump a pair of 0.5 mm thick beta-barium borate crystals, whose axes are oriented at right angles with respect to each other [13]. Each crystal produces Type-I spontaneous parametric down conversion, with signal and idler beams making angles of 3° from the pump. The relative amplitudes and phases of the two polarization components of the pump beam are controlled by a half-wave plate and a birefringent plate. Ideally the state that we are trying to produce is the Bell state

$$\left|\Phi^+\right\rangle = \frac{1}{\sqrt{2}}\left(\left|HH\right\rangle + \left|VV\right\rangle\right). \tag{3}$$

However, our experimentally produced state is not pure. We model the state produced by our source as

$$\hat{\rho}_s(p_s) = p_s \left|\Phi^+\right\rangle\left\langle\Phi^+\right| + \frac{1-p_s}{2}\left(\left|HH\right\rangle\left\langle HH\right| + \left|VV\right\rangle\left\langle VV\right|\right). \tag{4}$$

This density operator represents our photons as being in the Bell state $\left|\Phi^+\right\rangle$ with probability $p_s$, and in an equal mixture of the states $\left|HH\right\rangle$ and $\left|VV\right\rangle$ with probability $1-p_s$. A state of this type is produced, for example, if there is some temporal walk-off between the wave packets for horizontal and vertical polarizations, which introduces a degree of distinguishability between them.

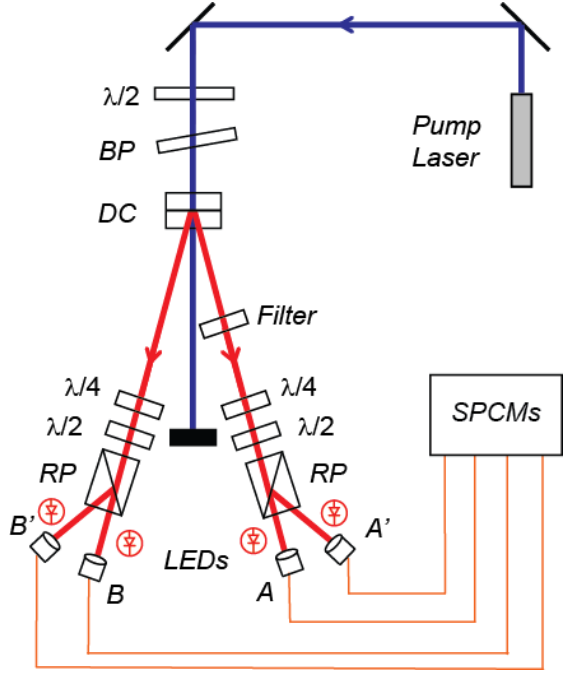

FIG. 1. The experimental configuration for performing two-qubit SPAM tomography. Here λ/2 denotes a half-wave plate, λ/4 denotes a quarter-wave plate, DC denotes down-conversion crystals, BP denotes a birefringent plate, RP denotes a Rochon polarizer, and SPCMs are single-photon-counting modules. The filter is an interference filter with a center wavelength of 810 nm, and a bandwidth of 10 nm.

In order to add an additional degree of freedom to our state preparation procedure, we illuminate each of our detectors with light from a light-emitting diode (LED). Adjusting the current in each LED provides a controllable amount of unpolarized, incoherent background to the detected state, and we model this state as

$$\hat{\rho}_w(p_s, p_w) = p_w \hat{\rho}_s(p_s) + \frac{1-p_w}{4}\hat{1} \ . \tag{5}$$

This is a state that is in a mixture of state $\hat{\rho}_s(p_s)$ with probability $p_w$, and a state of purely random polarization with probability $1-p_w$. In the limit that $p_s = 1$ the states in Eq. (5) are Werner states, and we describe the states $\hat{\rho}_w(p_s, p_w)$ as approximating Werner states [9, 14]. The states $\hat{\rho}_w(p_s, p_w)$ have a Horodecki $M$ parameter that is given by

$$M = p_w^2 + p_s^2 p_w^2. \tag{6}$$

**B. Experimental Results**

Coincidences are measured with a commercial time-to-digital converter; we use a coincidence resolution of 3.2 ns. We do not correct the data for accidental coincidences.

First we characterize our source with no background illumination from the LEDs. We measure a CHSH parameter of $S = 2.586 \pm 0.014$ (the uncertainty is the standard deviation of 10 independent measurements), which violates local realism by 40 standard deviations and indicates that the source produces entangled-photon pairs. We also perform traditional quantum-state tomography (QST) on this source to determine its density operator $\hat{\rho}$ [15]. We find that $\text{Tr}(\hat{\rho}^2) = 0.866$; this is a measure of the purity of the state. We then fit our measured $\hat{\rho}$ to the state $\hat{\rho}_w(p_s, p_w)$ of Eq. (5), using $p_s$ and $p_w$ as fit parameters. The best fit yields $p_s = 0.866$ and $p_w = 1.000$, which produces a fidelity between $\hat{\rho}_w(p_s, p_w)$ and $\hat{\rho}$ of $F = 0.985$. These parameters indicate that our source produces a state that is well described by Eq. (4). Using $\hat{\rho}$ and the method described in Sec. II.B we find $M = 1.726$, which yields $S_{\max} = 2.628$. This is consistent with our measured value.

Next we turn on the LEDs to create an approximate Werner state, and perform measurements in which Bob does not cheat. Alice performs measurements with 4 different settings of her quarter- and half-wave plate angles [(0,0), (π/4,π/8), (π/4,0), (π/8,π/16)]. Bob also uses 4 different settings for his wave plates [(π/8,π/16), (-π/8,-π/16), (π/4,0), (π/4,π/8)]. After Alice and Bob complete one trial of these 16 measurements they repeat them; in total they perform 10 trials. We estimate $\hat{\rho}$ by performing QST on all the data; the best fit $\hat{\rho}_w(p_s, p_w)$ has parameters $p_s = 0.928$ and $p_w = 0.628$, and a fidelity with $\hat{\rho}$ of $F = 0.997$. Furthermore, we find $M = 0.749$, which yields $S_{max} = 1.731$; this indicates that the state is incapable of violating the CHSH inequality. Alice's and Bob's first two measurement settings are those needed to measure the CHSH parameter. Using the data from these settings we find $S = 1.710 \pm 0.017$; as expected, this does not violate the CHSH inequality.

Even though this state does not violate the CHSH inequality, it is still entangled. The negativity $N$ of $\hat{\rho}$ is defined as

$$N = 2\sum_k \max(0, -\lambda_k) , \qquad (7)$$

where the $\lambda_k$'s are the eigenvalues of the partial transpose of $\hat{\rho}$ [16, 17]. For two-qubit states the negativity is 0 if the state is separable and 1 if it is maximally entangled. For our measured state we find $N = 0.397$. This value exceeds the lower bound on $N$ determined from our measured value of $S$ [18]

$$N \geq \frac{S}{\sqrt{2}} - 1 = 0.209 .$$

Finally, the 16 measurements of each trial are embedded in the 6x6 matrix $\overline{E}$, and the partial determinant $\Delta(\overline{E})$ is computed [8]. The 10 trials allow us to determine the statistics of $\Delta(\overline{E})$. In Fig. 2 we show the mean and standard deviation of $\Delta(\overline{E}) - \overline{1}$, and the ratio of the absolute value of these two quantities. By examining the absolute value of the ratio of the mean to the standard deviation [Fig. 2(c)] we find that all of the matrix elements of $\Delta(\bar{E}) - \bar{1}$ are 0 to within one standard deviation, which indicates that no correlated SPAM errors were detected. This agrees with our expectations, as Bob is not cheating and hence should not introduce any false correlations.

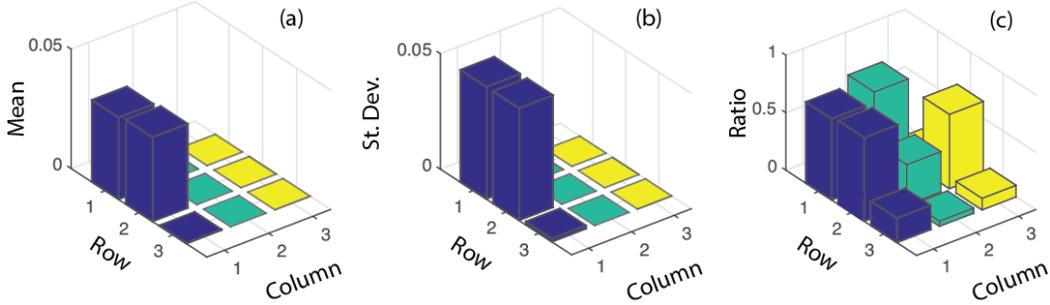

FIG. 2. (a) the mean of $\Delta(\bar{E}) - \bar{1}$, (b) the standard deviation of $\Delta(\bar{E}) - \bar{1}$, and (c) the absolute value of the ratio of these two quantities (mean divided by standard deviation) for the situation in which Bob does not cheat.

Now Alice and Bob make measurements using this same source, which is incapable of violating the CHSH inequality, but Bob cheats. He uses prior knowledge of Alice's measurement settings to adjust his own settings as follows. When Alice's wave plates are set to (0,0) he modifies his first two measurements: (π/8,π/16) → (0,0), (-π/8,-π/16) → (0,0). When Alice's wave plates are set to (π/4,π/8) he again modifies his first two measurements: (π/8,π/16) → (π/4,π/8), (-π/8,-π/16) → (-π/4,-π/8). Alice doesn't know that Bob has done this. With Bob's modified settings the measurements yield $S = 2.447 \pm 0.014$, which appears to violate the CHSH inequality by 32 standard deviations.

However, since Alice is suspicious of Bob she performs loop SPAM tomography on the full set of data; the results are shown in Fig. 3. We find that several matrix elements of $\Delta(\bar{E}) - \bar{1}$ differ from 0 by over 6 standard deviations [Fig. 3(c)]. This indicates that false correlations are present and that Bob's treachery has been detected.

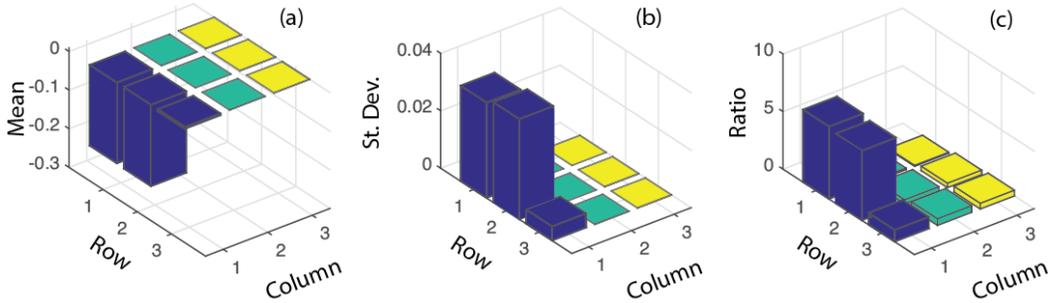

FIG. 3. (a) the mean of $\Delta(\bar{E}) - \bar{1}$, (b) the standard deviation of $\Delta(\bar{E}) - \bar{1}$, and (c) the absolute value of the ratio of these two quantities (mean divided by standard deviation) for the situation in which Bob cheats.

## IV. DISCUSSION AND CONCLUSIONS

A measurement that finds a value for the CHSH parameter of $S > 2$ indicates a violation of local-realism. Above we described an experiment with a source that was incapable of properly violating the CHSH inequality, but appeared to violate it. This was possible because Bob arranged to have prior knowledge of Alice's detector settings, and then used this knowledge to modify his own measurement settings. This allowed him to increase the apparent degree of correlation between their measurements [19]. Had Alice switched her measurement settings randomly, on a timescale faster than the speed of light travel time from her detector to and Bob's, then Bob would not have been able to manipulate measurements so as to yield $S > 2$.

Even though Alice did not prevent Bob from cheating, she was able to uncover his trickery by using loop SPAM tomography to detect the false correlations that he introduced. This is possible because loop SPAM tomography works without information about either the source state or the measurements, so Alice doesn't have to trust what Bob reports to her. The only information that Alice needs to perform loop SPAM tomography is the dimensions of the Hilbert spaces that describe the state of the source and the POVMs of the detectors.

Because it works in a device-independent manner, with a minimum of assumptions, loop SPAM tomography promises to be a useful tool for detecting correlated errors in a wide variety of quantum information processing systems.

**ACKNOWLEDGEMENTS**

M.E.F., G.K.J. and M.B. acknowledge the support of the National Science Foundation (PHY-1719390); they also acknowledge support from the Whitman College Louis B. Perry Summer Research Endowment and the Parents Student-Faculty Research fund. S.J.v.E. was supported in part by US Army Research Office under Contract No. W911NF-14-C-0048.

————————